\documentclass[aps,prl,twocolumn,amsmath,amssymb]{revtex4-2}

\usepackage{graphicx}
\usepackage{bm}
\usepackage{physics}
\usepackage{hyperref}

\begin{document}

\title{Relativity of Observation: Operational Intensive Variables in Nonequilibrium Thermodynamics}
\author{Akihisa Ichiki}
\email{ichiki@fukuoka-u.ac.jp}
\affiliation{Department of Applied Mathematics, Faculty of Science, Fukuoka University, 8-19-1, Nanakuma, Jonan-ku, Fukuoka City, 814-0180, Japan}

\date{\today}

\begin{abstract}
We formulate nonequilibrium thermodynamics in which intensive variables acquire operational meaning through measurement protocols consistent with local reciprocity. Using physical equilibrium as a reference, conjugate observables are constructed by continuously adjusting devices along the local tangent space of the statistical manifold. In this relativity of observation, Onsager reciprocity holds locally, allowing inference-based Lagrange multipliers to be directly measured. This provides a systematic method to extend operational definitions of intensive variables to nonequilibrium states, highlighting their context-dependent nature and offering a concrete experimental strategy.
\end{abstract}

\maketitle

\paragraph{Introduction.}
While Onsager reciprocity~\cite{onsager1931reciprocal} is often regarded as a consequence of microscopic reversibility near equilibrium, its role in nonequilibrium settings remains ambiguous. In particular, it is unclear whether reciprocity should be interpreted as a fundamental symmetry of dynamics, or rather as a property of observation and response defined relative to a chosen reference state. This ambiguity becomes crucial when intensive variables are introduced through inference~\cite{jaynes1957information, caticha2011entropic, caticha2008lectures}, where conjugate parameters arise mathematically but lack an a priori operational interpretation.

In this work, we adopt the latter viewpoint and reinterpret Onsager reciprocity as a local consistency condition for measurements. By treating macroscopic constraints as coordinates on a statistical manifold of inferred states~\cite{amari2000methods}, we show that reciprocity characterizes a locally flat observational frame, analogous to a local inertial frame in general relativity~\cite{thorne2000gravitation}. Our aim is not to develop a full information-geometric theory, but to identify the minimal structure required for intensive variables to become directly measurable through continuous, path-independent tuning of observation protocols.

\paragraph{Inference framework and conjugate observables.}
Consider a system with microscopic configurations $\omega$ and reference distribution $Q(\omega)$. Let $M_i(\omega)$ denote observable extensive variables with measured values $\mathbf{m}=(m_1, \cdots, m_n)$. The inference-based optimal distribution is obtained via the minimization of the relative entropy~\cite{jaynes2003probability, cramer1946mathematical}
\begin{equation}
P_\mathbf{m}^*=\displaystyle\arg\min_P D(P\|Q)
\end{equation}
subject to the constraints $\langle M_i\rangle_{P_\mathbf{m}^*}=m_i$, where $D(P\|Q)$ denotes relative entropy (Kullback--Leibler divergence), and $\langle\cdot\rangle_P$ is an expectation value with respect to the distribution $P$. The resulting inferred measure is
\begin{equation}
P_\mathbf{m}^*(\omega)=\frac{1}{Z(\bm{\lambda})}Q(\omega)\displaystyle\exp\left[-\sum_i\lambda_i M_i(\omega)\right],
\end{equation}
where $\lambda_i$ is a Lagrange multiplier corresponding to the measurement $M_i$, and $Z(\bm{\lambda})$ is a normalization constant. Accordingly, the Lagrange multiplier $\lambda_i$, which plays a role of intensive variable conjugate to $m_i$~\cite{jaynes1957information}, is obtained as an inferred quantity as
\begin{equation}
\lambda_i(\mathbf{m})=-\frac{\partial}{\partial m_i}D(P_\mathbf{m}^*\|Q).
\end{equation}

To assign an operational meaning~\cite{callen} to the conjugate variables $\bm{\lambda}$, we require that they be associated with measurements acting on the same microscopic configurations $\omega$ as the observables $\mathbf{M}(\omega)$. Since the family of optimal distributions $P^*_\mathbf{m}(\omega)$ is parametrized by the observable values $\mathbf{m}$, this requirement naturally leads us to consider how probability is redistributed in configuration space as $\mathbf{m}$ are varied. We therefore impose the existence of probability currents $J_i^\mathbf{m}(\omega)$ such that the change of the distribution with respect to $m_i$ satisfies an equation of continuity,
\begin{equation}
\partial_{m_i} P^*_\mathbf{m}(\omega)=-\partial_\omega J_i^\mathbf{m}(\omega).
\end{equation}

For normalized distributions, the continuity condition is consistent provided
\begin{equation}
\int d\omega\partial_{m_i} P^*_\mathbf{m}(\omega)=0,
\end{equation}
which is automatically satisfied since $\int d\omega P^*_\mathbf{m}(\omega)=1$ for all $\mathbf{m}$. Consequently, a set of currents $\{J_i^\mathbf{m}(\omega)\}_{i=1}^n$ solving the equation of continuity always exists, although it is not unique. This non-uniqueness reflects a gauge freedom in the choice of measurement protocol.

Now we consider observations $\Lambda_i^\mathbf{m}(\omega)$, which represent direct measurements of the inferred Lagrange multipliers: $\langle \Lambda_i^\mathbf{m}\rangle_{P_\mathbf{m}^*}=\lambda_i$. Such measurements should satisfy
\begin{equation}
\int\Lambda_i^\mathbf{m}(\omega)P_\mathbf{m}^*(\omega)d\omega= -\int\ln\frac{P_\mathbf{m}^*(\omega)}{Q(\omega)}\partial_{m_i} P_\mathbf{m}^*(\omega)d\omega.
\end{equation}
Then employing the equation of continuity yields
\begin{equation}
\Lambda_i^\mathbf{m}(\omega)=-\frac{J_i^\mathbf{m}(\omega)}{P_\mathbf{m}^*(\omega)}\partial_\omega\ln\frac{P_\mathbf{m}^*(\omega)}{Q(\omega)},
\end{equation}
if an appropriate (natural or periodic) boundary condition is applied.

At this stage, the observable $\bm{\Lambda}^\mathbf{m}(\omega)$ should be understood as an explicit construction demonstrating that the Lagrange multiplier associated with a constraint can, in principle, be given an operational measurement interpretation. The existence of a probability current $J_i^\mathbf{m}(\omega)$ satisfying the equation of continuity guarantees that such a conjugate observable can always be defined.

It is important to note that the observables $\bm{\Lambda}^\mathbf{m}(\omega)$ depend explicitly on the constraint values $\mathbf{m}$. This dependence does not indicate a failure of observability, but rather reflects the context-dependent nature of measurement: once a constraint is fixed, the corresponding conjugate observables are uniquely determined up to gauge degrees of freedom. In this sense, $\bm{\Lambda}^\mathbf{m}$ is not a set of universal measurement devices, but a set of observables defined relative to a specified inference context.

This result establishes that the intensive variables $\bm{\lambda}^\mathbf{m}$ acquire an operational meaning once probability currents consistent with the continuity condition is specified. Different choices of $\mathbf{J}^\mathbf{m}$ correspond to different measurement implementations, while yielding the same inferred distribution $P_\mathbf{m}^*$. In this sense, the observability of $\bm{\lambda}^\mathbf{m}$ is not a property of the distribution alone, but of the pair $(P_\mathbf{m}^*, \mathbf{J}^\mathbf{m})$.

\paragraph{Onsager reciprocity as a result of local flatness}
In a mult-constraint setting, the consistency condition $\partial_{m_i}\partial_{m_j}P_\mathbf{m}^*(\omega)=\partial_{m_j}\partial_{m_i}P_\mathbf{m}^*$ yields
\begin{equation}
\partial_{m_i}J_j^\mathbf{m}-\partial_{m_j}J_i^\mathbf{m}=0,
\label{eq:Jcommute}
\end{equation}
that is interpreted, through a linear response relation $J_i^\mathbf{m} = \sum_j L_{ij}m_j$ around $\mathbf{m}=\bm{0}$, as Onsager reciprocal relations~\cite{kubo1966fluctuation}.

From the viewpoint of information geometry, in the local regime where variations of $\mathbf{m}$ are restricted to the tangent space of the statistical manifold, the currents respond linearly to changes in constraints. The symmetry \eqref{eq:Jcommute} therefore plays the role of Onsager reciprocity, derived here not from microscopic reversibility but from the integrability of inference.

The family of inferred distributions generated by varying the constraint values $\mathbf{m}$ forms a smooth manifold embedded in the space of probability measures. Each point on this manifold corresponds to a macroscopic state specified by observable values, while nearby points represent infinitesimal changes of inference under perturbed constraints. The coordinates $\mathbf{m}$ thus parametrize a statistical manifold whose local structure is entirely determined by the inference procedure, independently of any underlying microscopic dynamics.

In general, this manifold is not globally flat: the response to changes in constraints depends on the reference state, and no single coordinate system renders all responses symmetric. Nevertheless, in the neighborhood of a given state $P_\mathbf{m}^*$, one may restrict attention to the tangent space spanned by admissible variations $\partial_{m_i}P_\mathbf{m}^*$. It is within this local regime that notions of conjugate variables and reciprocal response acquire an operational meaning. Accordingly, the observations for intensive variables explicitly depend on $\mathbf{m}$. This means that the intensive variables $\bm{\lambda}$ become observable only when the intensive measurements $\bm{\Lambda}^\mathbf{m}(\omega)$ are functions on the tangent space at $\mathbf{m}$. In such a frame, the probability currents $J_i^\mathbf{m}$ compatible with the equation of continuity always exist, and thus the reciprocal relation always holds.

In terms of mathematics, the quantity $\partial_{m_i}J^\mathbf{m}_j-\partial_{m_j}J^\mathbf{m}_i$ can be identified with the curvature of the effective connection on the statistical manifold generated by variations of the constraints. Onsager reciprocity corresponds to the vanishing of this curvature, i.e., to local flatness in the space of inferred states.

\paragraph{Gauge fixing of intensive observations: Continuity and commutativity of $\bm{\Lambda}^\mathbf{m}$.}
Let us consider the infinitesimal variation in $\mathbf{m}$, $\mathbf{m}\to\mathbf{m}+\delta\mathbf{m}$. Accordingly, the optimal measure changes as
\begin{equation}
P_\mathbf{m}^*(\omega)\to P_\mathbf{m}^*(\omega)+\displaystyle\sum_j \delta m_j\partial_{m_j}P_\mathbf{m}^*(\omega)|_{\mathbf{m}}+\mathcal{O}(\left(\delta\mathbf{m})^2\right).
\end{equation}
On the other hand, the intensive measurements $\bm{\Lambda}^\mathbf{m}$ have gauge degrees of freedom in currents,
\begin{equation}
J_i^{\mathbf{m}}(\omega)\to J_i^{\mathbf{m}}(\omega)+\tilde{J}_i(\omega),\qquad \partial_\omega\tilde{J}_i(\omega)=0.
\end{equation}
Thus the intensive measurements $\bm{\Lambda}^\mathbf{m}$ are not uniquely determined, but are observer-, or protocol-dependent in general.

Now we require that the intensive currents $J_i^\mathbf{m}$ vary continuously under infinitesimal changes of the macroscopic parameters $\mathbf{m}$. This requirement is expressed in equivalent form on the intensive observables as 
\begin{equation}
\displaystyle\lim_{\delta\mathbf{m}\to\bm{0}}\Lambda_i^{\mathbf{m}+\delta\mathbf{m}}(\omega)=\Lambda_i^\mathbf{m}(\omega).
\end{equation}
This regularity condition ensures that the differential form $\mathcal{J}^\mathbf{m} = \sum_i J_i^\mathbf{m}(\mathbf{m})\mathrm{d}m_i$ is locally exact, i.e., its value does not depend on the path taken in the space of $\mathbf{m}$. Equivalently, the integrability condition \eqref{eq:Jcommute} holds, implying the commutativity of mixed variations of the intensive observables. Under this condition, the set of the intensive observations $\bm{\Lambda}^\mathbf{m}$ is uniquely determined, which can be interpreted as a gauge fixing of intensive observables. This provides a minimal geometric structure sufficient to define a local inertial frame on the information manifold, without invoking further details of information geometry.

Starting from physical equilibrium as the reference $Q$, small deviations $\delta\mathbf{m}$ allow reciprocity conditions \eqref{eq:Jcommute} to be satisfied, ensuring uniqueness and continuity of $\Lambda^\mathbf{m}_i$. This defines a locally consistent measurement protocol where inferred variables $\lambda_i$ are directly observable.

\paragraph{Experimental implementation.}
To experimentally acquire intensive variables compatible with inference framework, we have to tune our observational setups. Tuning of such devices is acquired by the following protocol:
\begin{enumerate}
\item Reference calibration: Measure intensive variables $\lambda_i$ at physical equilibrium $\mathbf{m}=\mathbf{m}^0$. \item Incremental perturbation: Vary constraints $m_i\to m_i+\delta m_i$ within the linear regime.
\item Local adjustment: Tune measurement devices so that the reciprocity is satisfied.
\item Continuous mapping: Repeat steps 2--3 to span desired nonequilibrium states, ensuring a continuous, path-independent definition of $\bm{\Lambda}^\mathbf{m}$.
\item Measurement: Use the tuned devices $\bm{\Lambda}^\mathbf{m}$ to experimentally access inferred conjugate variables $\bm{\lambda}(\mathbf{m})$.
\end{enumerate}
Throughout these steps, the setups of extensive measurements $\mathbf{M}(\omega)$ are fixed. This protocol renders observer-independent measurements, starting from the equilibrium state as a reference $Q$.

\paragraph{Discussion.}
The central message of this work is that nonequilibrium thermodynamics admits a formulation closely analogous to general relativity, not at the level of microscopic dynamics, but at the level of observation. In general relativity, the equivalence principle asserts that locally, or within an appropriate inertial frame, gravitational effects can be eliminated and the laws of physics reduce to those of special relativity. Here, we show that an analogous principle holds for inference-based thermodynamics: although global reciprocity generally fails far from equilibrium, one can always construct a local observational frame in which Onsager reciprocity holds.

In our framework, the statistical manifold generated by inferred distributions plays the role of spacetime~\cite{ruppeiner1995riemannian}, while the macroscopic parameters $\mathbf{m}$ serve as coordinates defined through observation. The probability currents $J_i^\mathbf{m}$ encode how inference responds to infinitesimal changes of constraints, and their integrability condition, $\partial_{m_i}J_j^\mathbf{m}-\partial_{m_j}J_i^\mathbf{m}=0$, characterizes local flatness of this manifold. Onsager reciprocity is thus reinterpreted as a geometric statement: it holds precisely in those observational frames that act as local inertial frames on the information manifold.

From this perspective, Onsager relations are no longer universal laws tied to microscopic reversibility, but operational probes that identify locally flat regions of the inference geometry. By tuning measurement protocols so that reciprocity is satisfied, one effectively fixes a gauge and constructs intensive observables $\Lambda_i^\mathbf{m}$ that directly measure the inferred Lagrange multipliers $\lambda_i$. This provides a concrete operational meaning to intensive variables even far from physical equilibrium.

It is noting that this viewpoint also clarifies the coexistence of curvature and reciprocity. While the statistical manifold may be globally curved reflecting the irreversibility and context dependence of nonequilibrium inference, reciprocity always emerges locally in the tangent space around a given reference state. Just as gravity cannot be transformed away globally but disappears locally in a freely falling frame, entropy production~\cite{seifert2012stochastic} and asymmetric response need not vanish globally, yet Onsager symmetry can always be realized locally through appropriate observational tuning.

This interpretation leads naturally to an experimental strategy. Physical equilibrium serves as a reference point, analogous to a freely falling observer. Small deviations from equilibrium allow one to iteratively adjust measurement devices so that local reciprocity is maintained. In this way, Onsager relations function as experimentally accessible diagnostics for constructing local inertial frames in the space of inferred states. The operational measurement of intensive variables is therefore not an abstract inference procedure, but a realizable experimental protocol based on continuity, local linearity, and reciprocity.

Looking forward, this framework opens several directions. Connections to fluctuation relations, feedback-controlled systems, and quantum measurements suggest that the relativity of observation identified here may represent a general organizing principle for nonequilibrium physics. Some of details on such topics will be reported elsewhere.

\bibliographystyle{apsrev4-2}
\bibliography{measure_refs}

\end{document}